\documentclass[eqsecnum,twocolumn,tightenlines,floats,floatfix,prc,nofootinbib,preprintnumbers]{revtex4}%%

\def\bea{\begin{eqnarray}}
\def\eea{\end{eqnarray}}

\def\st#1{{\kern-4pt} \not\!#1}

\def\sp{\kern +3pt}
\def\sm{\kern -3pt}

\def\be{\begin{equation}}
\def\ee{\end{equation}}
\def\ba{\begin{eqnarray}}
\def\ea{\end{eqnarray}}

\usepackage{graphics}
\usepackage{graphicx}
\usepackage{epsf} 
\usepackage{amsmath}
\usepackage{amssymb}
\usepackage{slashed}
\usepackage{bm}

\usepackage{color}

\def\sfrac#1#2{{\textstyle \frac{#1}{#2}}}

\begin{document}

\phantom{0}
\vspace{-0.2in}
\hspace{5.5in}
%\parbox{1.5in}{ }

\preprint{ }

\vspace{-1in}%\parbox{1.5in}{ \vspace{-9.6in}}  % moves the preprint box down

\title
{\bf  
$\gamma^\ast N \to N^\ast(1520)$ form factors 
in the timelike regime}
\author{G.~Ramalho$^{1}$  and M.~T.~Pe\~na$^{2}$
\vspace{-0.1in}  }

\affiliation{$^1$International Institute of Physics, Federal 
University of Rio Grande do Norte, 
Campus Universit\'ario - Lagoa Nova, CP.~1613, 
Natal, Rio Grande do Norte 59078-970, Brazil
\vspace{-0.15in}}
\affiliation{$^2$Centro de F\'{i}sica Te\'orica e de Part\'{i}culas (CFTP),
Instituto Superior T\'ecnico (IST),
Universidade de Lisboa,
Avenida Rovisco Pais, 1049-001 Lisboa, Portugal}

\vspace{0.2in}
\date{\today}

\phantom{0}

\begin{abstract}
The covariant spectator quark model, tested before in 
a variety of electromagnetic baryon excitations,  is applied here
to the $\gamma^\ast N \to N^\ast(1520)$ reaction
in the timelike regime.
The transition form factors  are first  parametrized 
in the spacelike region
in terms of a valence quark core model together with 
a parametrization of the meson cloud contribution.
The form factor behavior in the timelike region
is then predicted, as well as the $N^\ast(1520) \to \gamma N$ 
decay width and the $N^\ast (1520)$ Dalitz decay, 
$N^\ast (1520) \to e^+ e^- N$.
Our results may help in the interpretation of dielectron production 
from elementary $pp$ collisions and from the new generation of HADES
results using a pion beam. In the $q^2=0$--1 GeV$^2$ range  
we conclude that the {\it QED approximation}
(a $q^2$ independent form factor model)
underestimates the electromagnetic coupling of the $N^\ast(1520)$
from 1 up to 2 orders of magnitude.
We conclude also that the $N^\ast (1520)$
and $\Delta(1232)$ Dalitz decay widths are comparable. 
\end{abstract}

%\phantom{0}
%\vspace{7.0in}
%\vspace{-6in}
\vspace*{0.9in}  % sets how far the title is below the preprint box
\maketitle

\section{Introduction}
\label{secIntro}

Measurements of dielectron production in elementary $pp$ 
collisions~\cite{MesonBeams,Faessler03,HADES14,HADES-1,HADES-2,Timelike,Timelike2} 
expand to the timelike region (\mbox{$q^2>0$}, where $q$ is the 
momentum transfer) the extraordinarily precise results 
obtained with the electron scattering 
CLAS $N^*$ program  at Jefferson Lab~\cite{MesonBeams,NSTAR,Aznauryan12a},
describing the electromagnetic structure of baryonic transitions 
in spacelike or  $q^2 <0$ region 
(program to be extended down to $- 12$ GeV$^2$, 
as well as up to $-0.05$ GeV$^2$).  
Although different, both experiments, with strong and electromagnetic 
probes, complement each other~\cite{MesonBeams}.

In the last years, $pp$ and quasi-free $pn$ reactions were combined for 
the determination of different medium effects, 
difficult to disentangle~\cite{Teis97,Bass98,Buss12}. 
The dominance of the $\Delta(1232)$ Dalitz decay for dielectron emission 
above the $\pi^0$ mass was confirmed by HADES, 
by the simultaneous measurement of one 
pion-production channels~\cite{HADES14,HADES-1}.
However, an excess of production with respect to the baryon 
and meson cocktail simulations was observed, an effect pointing 
to the contribution of 
low-lying resonances, as the 
$N^\ast(1520)$~\cite{HADES14,Ramstein16a,Przygoda16a}.
The important role of this state in dilepton decay reactions, 
as the $\gamma^\ast N \to e^+ e^- N$,
in the timelike region was also discussed in 
Refs.~\cite{Bratkovskaya99,Faessler03,Kaptari09,Weil12}.

To focus on this contribution, very recently, the High Acceptance 
Di-Electron Spectrometer (HADES) at GSI was combined with a pion beam 
for perfect and unique cold matter studies~\cite{HADES14,HADES-1,HADES-2}. 
The first pion beam results from 
HADES experiments~\cite{Ramstein16a,Przygoda16a} also
show in the dielectron emission channel a large peak in 
the number of events in a missing mass range of $0.9$--$1.04$ GeV$^2$, 
which is expected to be due to $N^\ast(1520)$ decay. 
The new experiments with HADES and a pion beam are an extraordinary opportunity  
to shed light on the behavior of the second and third resonance region, 
and the resonance form factors in the 
timelike region~\cite{HADES14,Ramstein16a,Przygoda16a}.
This motivates calculations, as the one reported here, 
for the extraction of different contributions 
to $e^+ e^-$ emission, as well as to revisit, 
in the electromagnetic couplings of baryons,
the extensively applied vector meson dominance principle and its validity.
To understand the structure of hadrons from first principles the 
nonperturbative character of QCD at low energies
and chiral symmetry have to be combined~\cite{NSTAR}.
In the two extreme regimes of QCD -- the 
low-energy regime where the energies are (much) smaller than 
a typical strong interaction scale and the high-energy regime 
where the energies are much higher than that scale -- 
well-established theoretical methods  
Chiral Perturbation Theory (ChPT) and 
perturbative QCD, respectively, apply~\cite{NSTAR}. 
However, in the intermediate-energy regime some degree 
of modeling is still required.  

Promising tools are Dyson-Schwinger-Faddeev functional methods and 
lattice QCD. Although progress in these fronts continues, they are developed
in Euclidean space and,  so far, are limited to the region below the 
$\rho$-meson pole~\cite{Eichmann16a}. 
At this stage, we take here a more phenomenological approach, 
based on the covariant spectator quark model.
This model is based on the covariant spectator theory 
(CST)~\cite{Gross}.
In a relatively successful 
and unifying way, our approach pictures a large variety of baryons 
as a superposition of a core of three valence quarks and meson cloud 
components~\cite{Timelike,Timelike2,NSTAR,Nucleon,Omega,Nucleon2,NDelta,NDeltaD,N1520,Roper,N1535,Delta1600,SQTM,Axial,Delta,NucleonDIS,OmegaGE2}.
After a series of applications of the model to the electromagnetic 
excitation of baryons in the spacelike regime,
we analyzed also the impact 
of the $\Delta(1232)$ resonance 
in the timelike reactions~\cite{Timelike,Timelike2}.

In this work we start with the quark model  described in 
Ref.~\cite{N1520} for the $N^\ast(1520)$ resonance 
and extend it to the region $q^2 > 0$.
In addition to the contribution from the 
bare core, we take also a meson cloud contribution.
This contribution is modeled within the lines of our previous study 
of the $\Delta(1232)$ 
in the timelike region, i.e., with the pion-photon coupling
parametrized by the pion form factor data~\cite{Timelike2}.

Three conclusions emerged in the context of our model: 
i) the $\gamma^\ast N \to N^\ast(1520)$ 
timelike transition form factors are dominated 
by the meson cloud contributions;
ii) in the range $q^2=0$--1 GeV$^2$ 
the constant form factor model 
(also known as {\it QED approximation}) usually taken in the literature
underestimates the electromagnetic coupling 
of the $N^\ast(1520)$ with consequences 
for the differential Dalitz decay width; 
iii)  in addition to the $\Delta(1232)$ resonance
the $N^\ast(1520)$ has a role
in dilepton decay reactions at intermediate energies.

This article is organized as follows:
In Sec.~\ref{secMethodology} we 
describe the methodology used to extend 
a valence quark  model fixed in the spacelike region to the timelike region.
Next, in Sec.~\ref{secDalitz}, 
we discuss the relation between the 
 $\gamma^\ast N \to N^\ast (1520)$ form factors 
and the formulas for the
photon and Dalitz decay widths of the $N^\ast (1520)$.
The formalism of the covariant spectator quark model used here
is presented briefly in Sec.~\ref{secSpectator}.
In Sec.~\ref{secFormFactors} we discuss 
the formulas used to calculate 
the  $\gamma^\ast N \to N^\ast (1520)$ form factors.
The results for the form factors in the timelike region  
and the $N^\ast (1520)$ decay widths are 
presented in Sec.~\ref{secResults}.
Outlook and conclusions 
are presented in Sec.~\ref{secConclusions}.

\section{Methodology}
\label{secMethodology}

In the covariant spectator quark model, the application 
of impulse approximation 
to the interaction of a photon with a baryon, seen as a 
three quark $qqq$ state,  
justifies that one integrates out the
relative internal momentum in the spectator 
diquark subsystem~\cite{Nucleon,Omega,Nucleon2}. 
After this internal momentum integration, 
in the process of the covariant integration over the global momentum 
of the interacting diquark one may  keep only the main contribution, 
which is originated by the  on-mass-shell pole of the diquark --- while
the  remaining quark that interacts with the photon  is taken to be 
off-mass-shell~\cite{Nucleon2}.
This last integration on the 
on-shell diquark internal momenta amounts to having the $qqq$ system 
as a quark-diquark system, and to treating the diquark with
an effective average mass $m_D$~\cite{Nucleon,Omega,Nucleon2}. 
It is also an ingredient of the model
that the electromagnetic quark current
is represented by a parametrization 
of vector meson dominance~\cite{Nucleon,Omega,Lattice,LatticeD}. 
In addition to the contributions from the core of
valence quarks, the covariant spectator quark model 
can include also a covariant parametrization of 
the meson cloud effects that are  
important in the low momentum transfer 
region and that depend on the baryons participating in 
\mbox{the reaction~\cite{Timelike,Timelike2,NDelta,NDeltaD,Delta1600,OctetFF,Transitions,Medium}.}

Here, the extension of the model to the timelike regime,
requires two important modifications:
\begin{itemize}
\item
The nucleon and the $N^\ast(1520)$ quark core wave functions 
have to be calculated in timelike kinematic 
conditions, depending on 
an arbitrary mass $W$ which 
can differ from the resonance mass, labeled $M_R$.
\item
The electromagnetic quark current has also to be extended 
to the timelike regime.
That is done by introducing finite  mass widths for 
the $\rho$ and $\omega$ mesons.
\end{itemize}

For the $\gamma^\ast N \to \Delta(1232)$ 
transition in the timelike region we have
 already found that the meson cloud contributions are important,
in comparison to the valence quark contributions~\cite{Timelike2}.
It is worthwhile now to test whether the same phenomena 
occurs for the $N^\ast(1520)$ resonance, which carries, in particular, 
a different isospin.
In our model the valence quark contributions for 
the magnetic and electric form factors 
vanish at the photon point ($q^2=0$) due to the orthogonality 
of the initial and final state wave functions~\cite{N1520}.
Other valence quark model estimate them as non-zero contributions
(a discussion can be found in Ref.~\cite{N1520}).
Since in our model, the valence quark contributions for 
the electric and magnetic transition form factors vanish at $q^2=0$,
their extension to the $q^2 > 0$ region
gives non-zero but small contributions for 
those transition form factors.
Nevertheless, our model can provide a good 
approximation for the $N^\ast(1520)$ resonance 
in the timelike region based on the 
meson cloud contributions, 
which dominate in the timelike region. 
Moreover, the form factors show a dependence on $q^2$
with consequences for the analysis of reactions in the timelike region, 
where the electromagnetic couplings are often fixed 
at their value at $q^2=0$ ({\it QED approximation}).

\section{$N^\ast(1520)$ Dalitz decay}
\label{secDalitz}

The $N^\ast(1520)$ resonance is a  $J^P= \frac{3}{2}^-$ state, with
isospin $I=\frac{1}{2}$.
The $N^\ast(1520)$ Dalitz decay into the nucleon can be expressed in terms 
of the decay width~\cite{Krivoruchenko02}
\ba
\Gamma_{\gamma^\ast N} (q,W)=
\frac{3 \alpha}{16} \frac{(W-M)^2}{M^2 W^3}
\sqrt{y_+ y_-} y_+ |G_T(q^2, W)|^2, \nonumber \\
\label{eqGamma1}
\ea 
where $q=\sqrt{q^2}$, 
$\alpha$ is the fine-structure constant,
\ba
y_\pm = (W \pm M)^2 -q^2,
\label{eqYpm}
\ea
and $|G_T(q^2,W)|^2$ is a combination 
of the electromagnetic transition form factors given by
\ba
& &
|G_T(q^2,W)|^2= \nonumber \\
& &
3 |G_M(q^2,W)|^2 + |G_E(q^2,W)|^2 + \frac{q^2}{2W^2} |G_C(q^2,W)|^2.
\nonumber \\
\label{eqGT}
\ea
In the previous equation 
$G_M$, $G_E$ and $G_C$ are, respectively, 
the magnetic dipole, electric and Coulomb
quadrupole form factors, which are
complex functions in the region $q^2> 0$.

The dilepton decay rate 
is obtained from the relation (\ref{eqGamma1}).
Using the compact notation 
$\Gamma \equiv \Gamma_{e^+ e^- N}$, 
one can calculate 
the dilepton decay rate~\cite{Krivoruchenko02,Faessler00} as
\ba
\Gamma_{e^+ e^- N}^\prime (q ,W) & \equiv & 
\frac{d \Gamma}{d q} (q,W) \nonumber \\
&=&  \frac{2 \alpha}{3 \pi q^3} 
(2 \mu^2 + q^2) \sqrt{1 - \frac{4 \mu^2}{q^2}}
\Gamma_{\gamma^\ast N}(q,W), \nonumber \\
\label{eqDalitzRate} 
\ea
where $\mu$ is the electron mass.

The Dalitz decay width is then determined 
by the integral of $\Gamma_{e^+ e^- N}^\prime (q,W)$ in
the kinematic region $4 \mu^2 \le q^2 \le (W-M)^2$:
\ba
\Gamma_{e^+ e^- N} (W) 
= \int_{2 \mu}^{W-M} \Gamma_{e^+ e^- N}^\prime (q ,W) \, dq.
\label{eqGammaDal}
\ea

\section{Covariant spectator quark model}
\label{secSpectator}

In the covariant spectator quark model 
the baryon wave functions are specified by the 
flavor, spin, orbital  angular momentum 
and radial excitations of the quark-diquark states that are
consistent with the baryon quantum number~\cite{NSTAR,Nucleon,Omega,OctetFF}.
The nucleon wave function $\Psi_N$ was obtained
in Ref.~\cite{Nucleon} and the 
wave function $\Psi_R$ of the 
resonance $N^\ast(1520)$  in Ref.~\cite{N1520}.
Those wave functions describe only the 
valence quark content of those baryons.

The constituent quark electromagnetic
current is written as the sum of a Dirac and a Pauli component, 
\ba
j_q^\mu(q) & =& \left( \frac{1}{6} f_{1+} + \frac{1}{2} f_{1-} \tau_3 
\right)  \gamma^\mu  + \nonumber  \\
& &     \left( \frac{1}{6} f_{2+} + \frac{1}{2} f_{2-} \tau_3 
\right) \frac{i \sigma^{\mu \nu} q_\nu}{2M},
\label{eqJq}
\ea
where $\tau_3$ is the Pauli matrix that acts on the   
(initial and final) baryon isospin states,
$M$ is the nucleon mass, and 
$f_{i\pm} (q^2)$ are the quark isoscalar/isovector form factors.
Those form factors will be parametrized with a form 
consistent with the vector meson dominance (VMD) mechanism.

For inelastic reactions we replace 
$\gamma^\mu \to \gamma^\mu - \frac{{\not  q} q^\mu}{q^2}$,
in order to ensure the conservation 
of the transition current.
This is is equivalent to the Landau  
prescription~\cite{Kelly98,Batiz98,Gilman02}.
The extra term restores current conservation but does not affect
the results of the observables~\cite{Kelly98}.

Once we know the wave functions for the nucleon $\Psi_N(P_-,k)$ 
and the resonance $\Psi_R(P_+,k)$, with momenta $P_-$ and $P_+$, 
respectively, and the diquark momentum $k$, we can calculate 
the transition current in a relativistic 
impulse approximation~\cite{Nucleon,Omega,Nucleon2}
\ba
J^\mu = 3 \sum_{\Gamma} \int_k 
\bar \Psi_R (P_+,k) j_q^\mu \Psi_N(P_-,k),
\label{eqJ1}
\ea
where $\Gamma$ represents the intermediate diquark polarizations,
and the integration symbol represents 
the covariant integration over the diquark 
on-shell momentum.
The factor 3 takes into account  the contributions 
of all of the quark pairs. The   
polarization indices are suppressed in the wave functions just for simplicity.
The current associated with the meson cloud 
will be parametrized separately and more phenomenologically. as
discussed later.
The two components of the current are conserved 
individually.

The definition (\ref{eqJ1}) for the electromagnetic current
is valid for the spacelike and timelike regions.
In the rest frame of the resonance (mass $W$), we may write
\ba
P_-=(E_N,-{\bf q}), \hspace{1cm} 
P_+=(W,{\bf 0}),
\label{eqPpPm}
\ea
where ${\bf q}$ is the photon three-momentum.
In that case the magnitude of the three-vector ${\bf q}$ 
corresponding to a photon of four-momentum $q$,
and the squared momentum $q^2$, is given by 
\ba
%|{\bf q}|^2 = \frac{[(W+M)^2-q^2][(W-M)^2-q^2]}{4W^2}.
|{\bf q}|^2 = \frac{y_+ y_-}{4W^2},
\ea
where $y_\pm$ is defined in Eq.~(\ref{eqYpm}).
In the case of a timelike photon ($q^2> 0$),
the last condition implies that physical 
photons (with $|{\bf q}|^2 \ge 0$) 
are defined only for $0 \le q^2 \le (W-M)^2$,
or $q^2 \ge (W+M)^2$.
As we are interested in resonance decay, the region near $q^2=0$
is the one where we will focus, and 
we will skip the discussion of the last case.
In conclusion, because both the nucleon and the resonance are taken on
their mass-shell the transition form factors 
for a transition between a nucleon of mass $M$
and a  resonance of mass $W$ are kinematically
restricted to the region 
$q^2 \le (W-M)^2$ in the timelike region. As the resonance mass $W$ 
grows larger,
the spanned momentum region increases.

\subsection{Quark form factors}

The valence quark form factors, included in the 
effective electromagnetic quark current (\ref{eqJq}) 
have a parametrization inspired in the
VMD mechanism that reads~\cite{Nucleon,Lattice,LatticeD}
\ba
& &f_{1\pm} (q^2)= \lambda_q 
+ (1-\lambda_q) \frac{m_{v\pm}^2}{m_{v\pm}^2- q^2} 
- c_\pm \frac{M_h^2 q^2}{(M_h^2-q^2)^2}, \nonumber \\
& &f_{2\pm} (q^2)= \kappa_\pm
\left\{ 
d_\pm
\frac{m_{v\pm}^2}{m_{v\pm}^2- q^2} +
(1- d_\pm) \frac{M_h^2}{M_h^2-q^2}\right\}.
\nonumber \\
\label{eqQFF}
\ea
Here, $m_{v\pm}$ represents light vector meson masses, 
$M_h$ is an effective heavy vector meson,
$\kappa_\pm$ indicates the quark anomalous magnetic moment,
$c_\pm,d_\pm$ are mixture coefficients, 
and $\lambda_q$ is a high-energy parameter 
related to the quark density number 
in the deep inelastic limit~\cite{Nucleon}.
For the isoscalar functions one has $m_{v+} = m_\omega$
($\omega$ mass) and for the isovector functions
one uses $m_{v-} = m_\rho$ ($\rho$ mass).
The term in $M_h = 2M$ simulates the effects 
of the heavier mesons and, therefore, all short range physics.

Specifically, we used the quark current 
parametrization of  model II  from Ref.~\cite{Nucleon}:
$\lambda_q=1.21$, 
$c_+=4.16$, $c_-= 1.16$, $d_+ = d_-=-0.686$, 
$\kappa_+= 1.639$ and $\kappa_-=1.833$.
The values were adjusted 
in order to describe the 
nucleon elastic form factor data in the spacelike region.
(The radial wave functions are described later).
Its behavior in that region was tested by taking it 
to the lattice QCD regime~\cite{Lattice,LatticeD},
and also to the nuclear medium~\cite{Medium}, 
both implemented with success.

However, some discussion is necessary for the timelike situation $q^2 >0$.
As seen from Eq.~(\ref{eqQFF}), singularities 
will appear when $q^2 = m_{v\pm}^2$. Physically they correspond to
the $\rho$ and $\omega$ poles.
Another singularity  appears from the pole at
 $q^2=M_h^2$, but only for very large $q^2$ ($\simeq 3.5$ GeV$^2$).
 The $M_h$-pole was introduced for 
phenomenological reasons to parametrize 
the short range physics in the spacelike region~\cite{Nucleon}.
For calculations with large $W$ (large $q^2$) 
the  $M_h$-pole has to be regularized as discussed 
in the Appendix. 

The spacelike parametrization of the quark current
in terms of the $\omega$ and $\rho$ poles
assumes that those particles are stable particles 
with zero decay width $\Gamma_v=0$.
In the extension of the quark form factors 
to the timelike regime we give them a width and use instead the substitution 
\ba
\frac{m_v^2}{m_v^2 -q^2} 
\to 
\frac{m_v^2}{m_v^2 -q^2 - i m_v \Gamma_v(q^2)},
\ea
where the index $v$ is used for either 
$\rho$ or $\omega$ as before. 
In r.h.s.~$\Gamma_v$ denotes the 
vector meson decay width function in terms of  $q^2$.

In the application to the 
$\Delta(1232) $ Dalitz decay~\cite{Timelike2}
only the $\rho$ pole was taken because
in the  $\gamma^\ast N \to \Delta$ transition 
only the isovector components contribute 
 (given by the functions $f_{i-}$).

The function $\Gamma_\rho(q^2)$ represents 
the $\rho \to 2 \pi$ decay width for 
a virtual $\rho$ with momentum $q^2$ 
\cite{Connell95,Muhlich}
\ba
\Gamma_{\rho} (q^2)=
\Gamma_{\rho}^0 \frac{m_\rho^2}{q^2}
\left(\frac{ q^2 - 4 m_\pi^2}{m_\rho^2- 4 m_\pi^2} \right)^{\frac{3}{2}}
\theta(q^2 - 4 m_\pi^2), 
\ea
where $\Gamma_{\rho}^0 = 0.149$ GeV.

\begin{figure}[t]
\vspace{.4cm}
\includegraphics[width=2.8in]{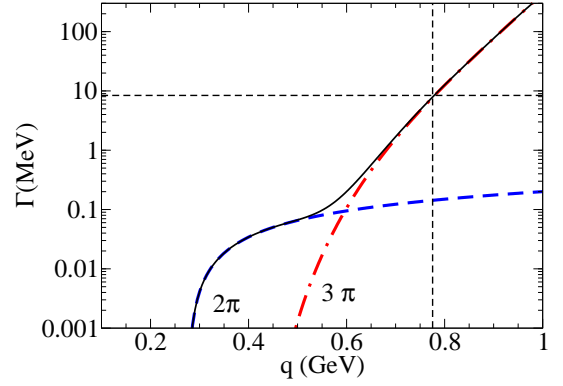}
\caption{\footnotesize
$\Gamma_\omega$ as a function of $q$.
The $2\pi$, $3\pi$ channels are indicated by the 
long-dashed and dotted-dashed lines respectively.
The solid line represents the sum of the two channels.
The short-dashed vertical and horizontal 
lines indicate the $\omega$ mass point 
and the $\omega $-physical width (8.4 MeV).}
%\vspace{-1cm}
\label{figGamma}
\end{figure}

For the application in this paper, however, 
we also have to include the $\omega$ pole.
To this end, 
the function $\Gamma_\omega(q^2)$ will include 
the decays $\omega \to 2 \pi$ (function $\Gamma_{2\pi}$)
and $\omega \to 3 \pi$ (function $\Gamma_{3\pi}$).
The case $\omega \to 3 \pi$ can be interpreted
as the process $\omega \to \rho \pi \to 3 \pi$, and therefore
we decomposed $\Gamma_\omega(q^2)$ into
\cite{Muhlich} 
\ba
\Gamma_\omega(q^2)= \Gamma_{2\pi} (q^2) +  \Gamma_{3\pi} (q^2),
\label{eqGammaOm}
\ea
The function $\Gamma_{2 \pi}$ can be represented as~\cite{Connell97,Muhlich}
\ba
\Gamma_{2\pi} (q^2)=
\Gamma_{2\pi}^0 \frac{m_\omega^2}{q^2}
\left(\frac{ q^2 - 4 m_\pi^2}{m_\omega^2- 4 m_\pi^2} \right)^{\frac{3}{2}}
\theta(q^2 - 4 m_\pi^2), 
\ea
where $\Gamma_{2\pi}^0 = 1.428 \times 10^{-4}$ GeV.
Note that $\Gamma_{2\pi}$ is similar 
to the function $\Gamma_\rho$ except for the constant $\Gamma_{2\pi}^0$ 
(about $10^3$ smaller) and the mass.
For the function $\Gamma_{3\pi}$ we use the 
result from Ref.~\cite{Muhlich}
\ba
\Gamma_{3\pi} (q^2)= 
\int_{9m_\pi^2}^{(q-m_\pi)^2} ds \; 
{\cal A}_\rho (s)
\bar \Gamma_{\omega \to \rho \pi} (q^2,s) ,
\ea
where $q=\sqrt{q^2}$, $s$ the mass of the virtual $\rho$ meson,
$\bar \Gamma_{\omega \to \rho \pi}$ is the decay width 
of $\omega$ to a $\pi$ and a virtual $\rho$
and ${\cal A}_\rho$ is the $\rho$-spectral function.
The
functions $\bar \Gamma_{\omega \to \rho \pi}$ and 
${\cal A}_\rho$ are~\cite{Muhlich}
\ba
\bar \Gamma_{\omega \to \rho \pi} (q^2,s) &=&
\frac{3}{4\pi} \left( \frac{g'}{m_\pi} \right)^2 
\left[
\frac{(q^2-s -m_\pi^2)^2- 4 s m_\pi^2}{4 q^2}
\right]^{\frac{5}{2}}  \nonumber \\
&& \times \theta(q^2- 9 m_\pi^2),
\ea
with $g'=10.63$ MeV  
and
\ba
{\cal A}_\rho (s)= \frac{\sqrt{s}}{\pi} 
%\frac{\Gamma_\rho(s)}{(s-m_\rho^2)^2 + s [\Gamma_\rho(s)]^2}. 
\frac{\Gamma_\rho(s)}{(s-m_\rho^2)^2 + s \Gamma_\rho^2(s)}. 
\ea
With this parametrization we obtain 
$\Gamma_\omega (m_\omega^2) \simeq \Gamma_{3 \pi} (m_\omega^2) = 7.6$ MeV,
which is consistent with the data.
Note that  the total width of the $\omega$
comprises the decays into 
$\gamma \pi, 2 \pi$ and $3 \pi$, and is 8.4 MeV.
The remaining contribution to the $\omega$ decay width comes from
the decay $\omega \to \gamma \pi^0$.
The $3\pi$ decay corresponds to a branching ratio of about 90\%.

The result of the calculation of $\Gamma_\omega$ 
as a function of $q$ is shown in Fig.~\ref{figGamma}.
Note in this figure the dominance of the $3\pi$ channel 
for $q > 0.55$ GeV.

\section{Form factors}
\label{secFormFactors}

\begin{figure}[t]
%\vspace{.2cm}
\includegraphics[width=2.8in]{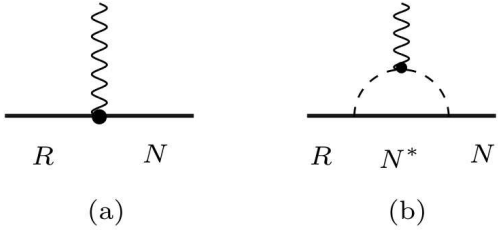}
\caption{\footnotesize
Electromagnetic interaction with  
the quark core (a) and with the meson cloud (b).
The intermediate $N^\ast$ is a 
octet baryon member (spin 1/2)
or a decuplet baryon member (spin 3/2).}
%\vspace{-1cm}
\label{figPionCloud}
\end{figure}

Although the  $\gamma^\ast N \to N^\ast(1520)$ transition 
is characterized by the three independent form factor functions
$G_M,G_E$ and $G_C$, in the study of the reaction in the 
spacelike regime~\cite{N1520} we concluded that
it is convenient to define 
an auxiliary form factor 
\ba 
\tilde G_4 = G_E + G_M,
\label{eqG4til}
\ea
because the valence quark contributions 
for $\tilde G_4$ are zero and a direct extraction of
the {\it pure}  meson cloud contribution from data 
 therefore arises naturally
(in the context of the covariant spectator quark model).
The valence quark contributions for the 
transition form factors are discussed next.

Separating the  valence quark contribution, 
represented in Fig.~\ref{figPionCloud}(a), 
from the meson cloud contribution, 
represented in Fig.~\ref{figPionCloud}(b),
one can decompose each of the three form factors as in Ref.~\cite{N1520}
\ba
& &
\hspace{-1.2cm}
G_M(q^2,W)= G_M^B (q^2,W) + G_M^\pi (q^2), \\
& &
\hspace{-1.2cm}
G_E(q^2,W)= -G_M^B (q^2,W) - G_M^\pi(q^2) 
+ \tilde G_4^\pi(q^2), \\
& &
\hspace{-1.2cm}
G_C(q^2,W)= G_C^B(q^2,W) + G_C^\pi(q^2),
\ea 
where $G_X^B(q^2,W)$, $X=M,E,C$ give the valence quark core contributions 
and $G_M^\pi, \tilde G_4^\pi$
and $G_C^\pi$ stand for the matching  meson cloud contributions.
The label $\pi$ is used instead of $MC$ 
(meson cloud) because we describe
those contributions 
in terms of the pion electromagnetic form factor (following what was done  for
the $\gamma^\ast N \to \Delta$ transition~\cite{N1520,Timelike2}).
The formulas for the  valence quark core
and meson cloud contributions will be given in the next sections.

Once the transition form factors are known,
the  helicity amplitudes can be calculated using
\ba
& &
A_{1/2} = \frac{1}{F} G_M + \frac{1}{4F} \tilde G_4, \\
& &
A_{3/2} = \frac{\sqrt{3}}{4F} \tilde G_4, \\
& &
S_{1/2} = \frac{1}{\sqrt{2} F} \frac{|{\bf q}|}{2 W} G_C, 
\ea
where
$F = \frac{1}{e} \frac{W}{|{\bf q}|} 
\sqrt{\frac{MK}{W}  \frac{y_-}{(W-M)^2}}$ 
with $K= \frac{W^2- M^2}{2W}$.

\subsection{Valence quark form factors}

The contributions of the valence quark core
to the form factors can be calculated and seen to have the general 
final form~\cite{Aznauryan12a,Devenish76}
\ba
G_M^B &=&
- {\cal R} \left[ (W-M)^2 -q^2 \right] \frac{G_1}{W}, 
\label{eqGM}\\
G_E^B &=&-
{\cal R} \left\{ 4 G_4 - 
\left[ (W-M)^2 -q^2 \right] \frac{G_1}{W} \right\}, \nonumber \\
& &
\label{eqGE} \\
G_C^B &=& - {\cal R} \left[
4W G_1 + (3 W^2+M^2-q^2)G_2 \right. \nonumber \\
& & 
\left. + 
2 (W^2-M^2 + q^2) G_3 \right],
\label{eqGC}
\ea
where $G_i$ ($i=1,2,3$) are three independent form factors and
$ {\cal R}=  \frac{1}{\sqrt{6}}\frac{M}{W-M}$.
The function $G_4$ was introduced for convenience.
Because of current conservation 
$G_4$ is given in terms of three independent $G_i$ ($i=1,2,3$) 
as~\cite{N1520}
\ba
G_4= (W- M) G_1 + \frac{1}{2}(W^2-M^2)G_2 + q^2 G_3. \nonumber \\
\label{eqG4}
\ea
By combining the results for
$G_E$ and $G_M$ one concludes that
the valence quark core contribution for $\tilde G_4$ 
is $\tilde G_4^B= G_E^B + G_M^B= - 4 {\cal R} G_4$.

The explicit calculation of the form factors 
requires the determination of the 
coefficients of the anti-symmetric ($A$)
and symmetric ($S$) components of the 
wave functions, in terms of the 
quark form factors  
\ba
j_i^A &=& \frac{1}{6} f_{i+} + \frac{1}{2} f_{i-} \tau_3, 
\label{eqJA}
\\
j_i^S  &=& 
\frac{1}{6}f_{i+} -\frac{1}{6} f_{i-} \tau_3.
\label{eqJS}
\ea
See Refs.~\cite{OctetFF,N1520} for more details. 

Then, the functions $G_i$ can be computed from 
the nucleon and resonance wave functions.
The results are~\cite{N1520}
\ba
G_1&=& -\frac{3}{2\sqrt{2} |{\bf q}|} \nonumber \\
& & \times
\left[
 \left( j_1^A + \frac{1}{3} j_1^S\right)+ 
\frac{W+M}{2M}\left( j_2^A + \frac{1}{3} j_2^S\right) 
\right] {\cal I}, \nonumber \\
& & \label{eqG1_P1}\\
G_2 &=& 
 \frac{3}{2\sqrt{2} M |{\bf q}|}  
\nonumber \\
& & \times
\left[
j_2^A 
+ \frac{1}{3}  \frac{1- 3 \tau}{1 + \tau}   j_2^S
+ 
\frac{4}{3} \frac{2 M}{W +M} \frac{1}{1 + \tau} j_1^S 
\right] {\cal I}, \nonumber \\
& &  \\
G_3&=& \frac{3}{2\sqrt{2} |{\bf q}|} 
\frac{W-M}{q^2}  
\nonumber \\
& &
\times \left[
j_1^A + \frac{1}{3}\frac{\tau-3}{1+ \tau} j_1^S 
+ \frac{4}{3} \frac{W +M}{2M} \frac{\tau}{1+ \tau} j_2^S
\right] {\cal I}, \nonumber \\
\label{eqG3_P1}
\ea 
where $\tau = -\frac{q^2}{(W+M)^2}$, and 
\ba
{\cal I}= - \int_k  
\frac{(\varepsilon_{0 P_+} \cdot \tilde k)}{\sqrt{- \tilde k^2}} 
\psi_{R}(P_+,k) 
\psi_N(P_-,k).
\label{eqInt}
\ea
The functions $\psi_N$ and $\psi_R$  in the formulas above
are the nucleon and  
resonance radial wave functions,  respectively; 
$\varepsilon_{0 P_+}$ is the spin-1 polarization 
vector and 
$\tilde k = k - \frac{P + \cdot k}{W^2} P_+$~\cite{N1520}.
The previous integral is calculated 
in the resonance rest frame using 
the conditions given by Eqs.~(\ref{eqPpPm}).

From Eq.~(\ref{eqG4}) one concludes, as mentioned before, 
that $G_4 \equiv 0$, implying that $G_E^B=-G_M^B$,
and motivating the use 
of $\tilde G_4$ to extract direct information on
the meson cloud since it, alone, contributes to $\tilde G_4$.

The consequence of the gauge invariant correction to Eq.~(\ref{eqJq})
is that in the Dirac part of the current, 
$G_3$  is determined from $G_1$ and $G_2$, 
and $G_3$ becomes non-zero at $q^2=0$~\cite{N1520}.
Importantly, the Dirac contribution for
$G_3$ is responsible for the non-vanishing value of $G_C$ at $q^2=0$.
This is similar to the $\Delta(1232)$ case addressed in Ref.~\cite{NDeltaD}.

The radial wave functions $\psi_N$ and $\psi_R$ 
are parametrized phenomenologically as in  Ref.~\cite{N1520} 
for the $N^\ast(1520)$ resonance, and as in
Ref.~\cite{Nucleon} for the nucleon.
Those functions depend on $P \cdot k$,
where $P$ is the momentum of the baryon.
More specifically, those functions 
can be expressed as functions of the dimensionless 
variable $\chi$ which in the nonrelativistic limit 
becomes proportional to ${\bf k}^2$~\cite{Omega}
and is defined as
\ba
\chi_B = \frac{(M_B-m_D)^2 - (P-k)^2}{M_B m_D},
\ea
where $M_B$ is the mass of the baryon.

The nucleon radial wave function is
represented as~\cite{Nucleon}
\ba
\psi_N(P,k) = \frac{N_0}{m_D (\beta_1 + \chi_N) (\beta_2 + \chi_N)},
\ea
where $\beta_1$ and $\beta_2$ are two momentum range 
parameters and $N_0$ is the normalization constant.
We choose $\beta_2 > \beta_1$, therefore 
$\beta_2$ regulates the long range behavior 
in the configuration space.
In the numerical calculations we used $\beta_1= 0.049$ and $\beta_2= 0.717$~\cite{Nucleon}.

For the $N^\ast (1520)$ state we used~\cite{N1520}
\ba
\psi_R (P,k)= 
\frac{N_1}{m_D (\beta_2 + \chi_R)}
\left[
\frac{1}{(\beta_1 + \chi_R)}
- 
\frac{\lambda_R}{(\beta_3 + \chi_R)}
\right],
\nonumber \\
\ea
where $N_1$ is the normalization constant,
$\beta_3$ is a new (short range) 
parameter, and $\lambda_R$ is a parameter 
determined by an orthogonality condition 
between the nucleon and the $N^\ast (1520)$ wave functions.
As in Ref.~\cite{N1520} we use here $\beta_3= 0.257$.
The orthogonality condition for wave functions of the nucleon 
and its excitation is given by 
${\cal I}(0)=0$, where ${\cal I}(Q^2)$ is defined by 
Eq.~(\ref{eqInt})~\cite{N1520}.

The parameters of the nucleon radial wave 
function were determined by the direct fit 
to the nucleon form factor data, 
in a model with no meson cloud~\cite{Nucleon}.
The parameters of the $N^\ast(1520)$ radial 
wave function were determined by 
the fit to the $\gamma^\ast N \to N^\ast(1520)$ 
data for $Q^2 =- q^2 > 1.5$ GeV$^2$,
a region where the meson cloud effects are 
expected to be very small~\cite{N1520}.

The radial wave functions $\psi_N$ and $\psi_R$ 
are normalized, by imposing the conditions
($B=N,R$)
\ba
\int_k |\psi_B (\bar P,k)|^2 =1, 
\label{eqPsiNorm}
\ea
where $\bar P=(M_B,0,0,0)$ is the baryon
momentum in the rest frame ($M_N$ represents the nucleon mass).
Equation~(\ref{eqPsiNorm}) ensures 
the right charge for each of the baryons $B$, obtained from the operator
$\sfrac{1}{2}(1+ \tau_3)$~\cite{Nucleon,N1520}.

\subsection{Meson cloud form factors}
\label{secMCFF}

The meson cloud form factors can be 
represented as in Ref.~\cite{N1520} 
\ba
\tilde G_4^\pi(q^2) &=& \lambda_\pi^{(4)}
 \left(\frac{\Lambda_4^2}{\Lambda_4^2- q^2}\right)^3
F_\pi (q^2) \; \tau_3, \label{eqG4pi}\\
G_M^\pi (q^2) &=& (1 - a_M q^2) \times \nonumber \\
& & 
\lambda_\pi^{M} \left(\frac{\Lambda_M^2}{\Lambda_M^2- q^2}\right)^3
F_\pi (q^2) \; \tau_3, 
\label{eqGMpi} \\
G_C^\pi (q^2) &=& \lambda_\pi^{C} \left(\frac{\Lambda_C^2}{
\Lambda_C^2 - q^2}\right)^3 
F_\pi (q^2) \; \tau_3,
\label{eqGCpi}
\ea
where $F_\pi(q^2)$ is a  parametrization of 
the pion electromagnetic form factors determined in Ref.~\cite{Timelike2}. 
Specifically, we use the form 
\ba
F_\pi (q^2) = \frac{\alpha}{\alpha - q^2  - 
\frac{1}{\pi} \beta q^2 \log \frac{q^2}{m_\pi^2} 
+ i \beta q^2},
\ea
where  $\alpha = 0.696$ GeV$^2$, $\beta= 0.178$, 
and $m_\pi$ is the  pion mass.

In our first work in Ref.~\cite{N1520} the meson cloud 
was different than the one that we are using here. 
The reason is that the meson model associated with the diagram 
Fig.~\ref{figPionCloud}(b) was, meanwhile, 
reparametrized in Ref.~\cite{Timelike2}
to fix the incorrect position of
the rho mass pole given by our first model, as well as by other 
popular parametrizations~\cite{Timelike2}.
In addition, we notice that in this new parametrization the 
$\gamma^\ast N \to \Delta$ transition pion cloud 
is directly connected to the pion 
electromagnetic form factor $F_\pi (q^2)$, which is well-established 
experimentally in the timelike region~\cite{Timelike2}.

\begin{figure}[t]
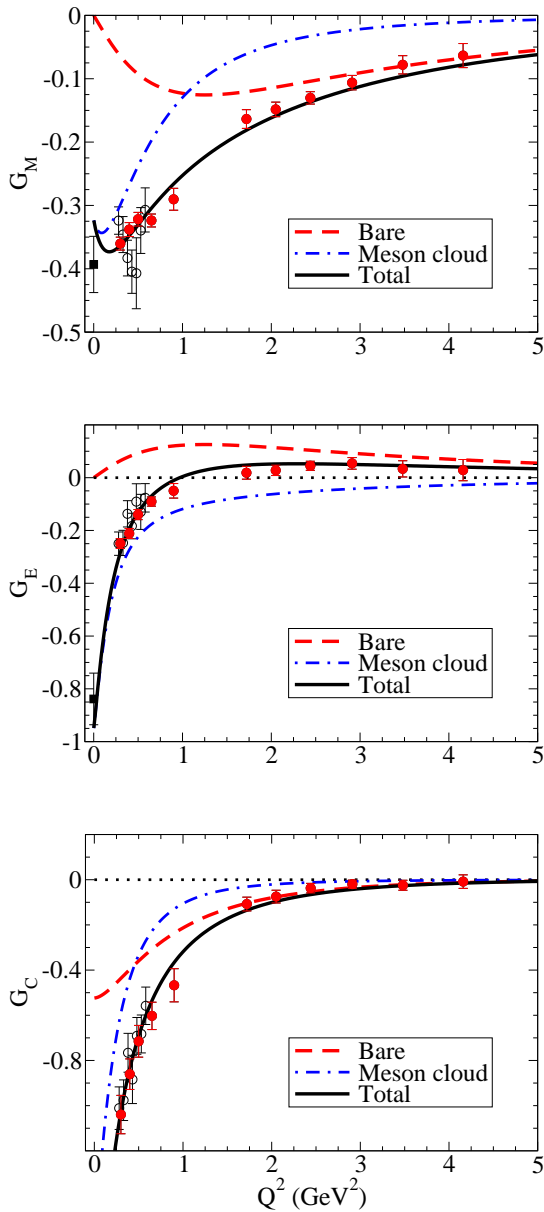

\vspace{.3cm}
\centerline{
\mbox{
\includegraphics[width=2.8in]{GM_mod4}
}}
\centerline{
\vspace{.5cm} }
\centerline{
\mbox{
\includegraphics[width=2.8in]{GE_mod4}
}}
\centerline{
\vspace{.5cm} }
\centerline{
\mbox{
\includegraphics[width=2.8in]{GC_mod4}
}}
\caption{\footnotesize{
Valence quark core plus meson cloud contributions to
the spacelike form factors as a function of $Q^2=-q^2$. 
Data from Ref.~\cite{Aznauryan09} (full circles),
Ref.~\cite{Mokeev12} (empty circles) and PDG~\cite{PDG} (square).}}
\label{figFF4}
\end{figure}

The parameters used
in the formulas (\ref {eqG4pi})-(\ref {eqGCpi}) were
determined by their fit to the 
$\gamma^\ast N \to N^\ast(1520)$ spacelike form factors, giving 
$a_M=5.531$ GeV$^{-2}$,
$\lambda_\pi^{(4)}= - 1.019$, 
$\lambda_\pi^M=-0.323$, 
$\lambda_\pi^C= -1.678$,
$\Lambda_4^2= 10.2$ GeV$^2$, 
$\Lambda_M^2= 1.241$ GeV$^2$ and 
$\Lambda_C^2= 1.263$ GeV$^2$.
The results are presented in Fig.~\ref{figFF4} as a function of $Q^2=-q^2$
and compared with the spacelike data~\cite{Aznauryan09,Mokeev12,PDG}.
Check Ref.~\cite{N1520} 
for a more detailed discussion of the data.
In the figure we also show  the  
the valence quark contributions (dashed line)
and the meson cloud contributions (dashed-dotted line)
based on the parametrizations described above.

In the Appendix, we discuss the technical aspects of the regularization
of the singularities appearing in the 
multipoles of Eqs.~(\ref{eqG4pi})-(\ref{eqGCpi}).

\section{Results}
\label{secResults}

We present in this section our predictions for the 
$\gamma^\ast N \to N^\ast(1520)$
transition form factors in the timelike region.
Using these results, we also calculate the $\gamma N$ and $e^+ e^- N$ 
decay widths.

\subsection{Form factors}

\begin{figure}[t]
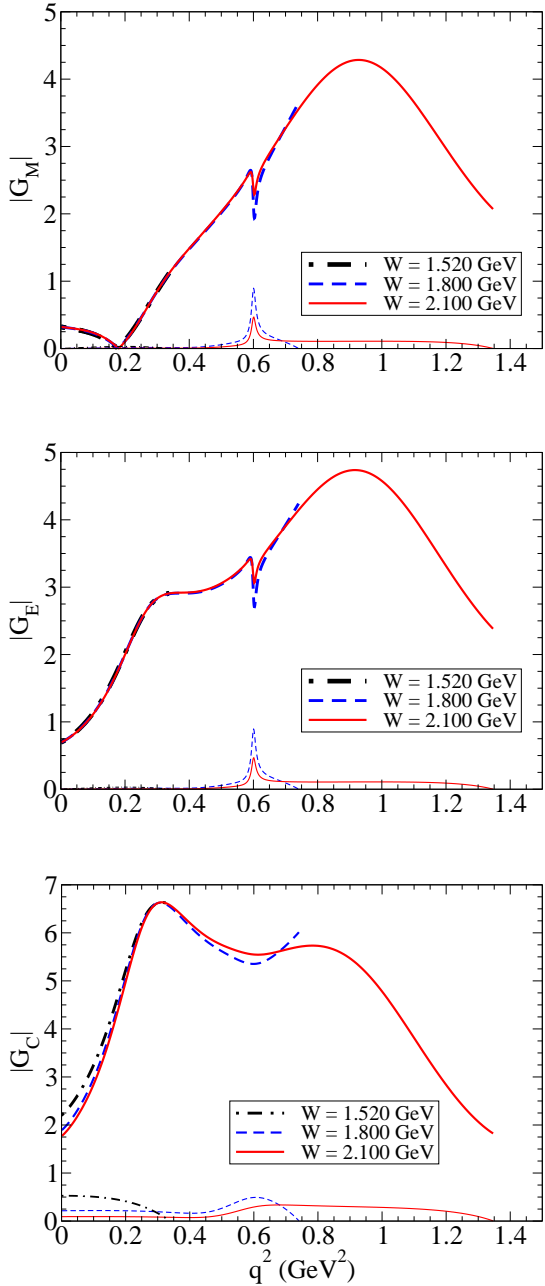

\vspace{.3cm}
\centerline{
\mbox{
\includegraphics[width=2.8in]{GMabs5b}
}}
\centerline{
\vspace{.5cm} }
\centerline{
\mbox{
\includegraphics[width=2.8in]{GEabs5b}
}}
\centerline{
\vspace{.4cm} }
\centerline{
\mbox{
\includegraphics[width=2.8in]{GCabs5b}
}}
\caption{\footnotesize{  
Absolute values of the form factors for 
$W=1.520, 1.800$ and 2.100 GeV.
In the calculations we use the new parametrization 
and the width $\Gamma_\omega(q^2)$ given by Eq.~(\ref{eqGammaOm}).
The thin lines represent the contribution from the core.
For the total result (thick lines) the lines for $W=1.520$ GeV coincide 
with the lines 
for $W=1.800$ and 2.100 GeV.
}}
\label{figGX2}
\end{figure}

The predictions for the absolute values of the 
form factors $G_M, G_E$ and $G_C$ in the timelike region
are presented in Fig.~\ref{figGX2}
for the cases $W=1.52,$ 1.8 and 2.1 GeV.
The valence quark core contributions are 
given 
by the thin lines. They stand very near the horizontal axis, and 
vanish in the upper limit $q^2 = (W-M)^2$ 
by kinematic constraints.
The same result was observed 
in the quadrupole form factors 
of the $\gamma^\ast N \to \Delta(1232)$ transition
for the physical case, when 
$W = M_\Delta \simeq 1.232$ GeV~\cite{DeltaSiegert2}. 
 
Figure~\ref{figGX2} shows that the meson cloud contribution
largely dominates.
Only near the $\omega$-pole ($q^2 \simeq 0.6$ GeV$^2$)
is there a significant contribution 
from the  quark core for the absolute value of the form factors
$G_M$ and $G_E$.
This effect is very concentrated near $q^2 \simeq m_\omega^2$ 
as a consequence of the small 
$\omega$ width, $\Gamma_\omega (m_\omega^2)$.

In $G_C$ the effect of the $\omega$ pole is not observed.
This is due to  
the cancellation of the isoscalar contributions 
to the form factor $G_C$.
This cancellation is obtained analytically  and can be confirmed by
%replacing 
substituting the form factors $G_1,G_2,G_3$ 
given by Eqs.~(\ref{eqG1_P1})-(\ref{eqG3_P1})
into the formula of Eq.~(\ref{eqGC}) for $G_C$.
One concludes that only the quark isovector form factors, 
$f_{1-}$ and $f_{2-}$, contribute to $G_C$.

From Fig.~\ref{figGX2}  one concludes that a fairly good description 
of the $\gamma^\ast N \to N^\ast(1520)$ transition can 
be obtained without the valence quark core contributions, which are very small.
The almost perfect coincidence, both for $G_M$ and $G_E$, of the lines
corresponding to different values of $W$  
is also a consequence of the dominance of the meson cloud component,
since only the valence part depends on $W$.
Only for $G_C$ can one distinguish 
a slight $W$ dependence, and this is evident because the 
valence quark contributions are nonzero when $q^2=0$.
The main role of the mass dependence $W$ in the behavior of the form factors 
is then to constrain them for $q^2 \le  (W-M)^2$.

Qualitatively, one can say that the form factors are enhanced around 
the origin up to a certain maximum value of $q^2$.
For $W=2.1$ GeV,
the magnitude of the form factors starts to decrease 
after $q^2 \approx 1$ GeV$^2$.
This effect is a consequence of the meson cloud parametrization
by Eqs.~(\ref{eqG4pi})-(\ref{eqGCpi}),
which includes a cutoff for $q^2$ around 1.2 GeV$^2$.
For larger values of $W$, the falloff 
of the form factors can also be observed.

\begin{figure}[t]
\vspace{.3cm}
\centerline{
\mbox{
\includegraphics[width=2.8in]{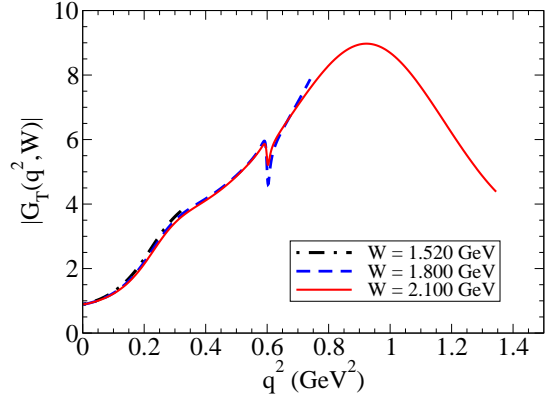}
}}
\caption{\footnotesize{
Effective contribution of the form factors 
$|G_T (q^2,W)|$ for 
$W=1.520, 1.800$ and 2.100 GeV.
}}
\label{figGT}
\end{figure}

The impact of the transition form factors 
in the timelike reactions is determined by 
the absolute value of  $|G_T (q^2,W)|$ given by Eq.~(\ref{eqGT}).
The results for  $|G_T (q^2,W)|$ 
for $W=1.52$, 1.8 and 2.1 GeV
are presented in Fig.~\ref{figGT}.
There is an increase of  $|G_T (q^2,W)|$ relative 
to its value at the origin
up to $q^2 \simeq 0.9$ GeV$^2$, for $W > 1.8$ GeV.
Above $q^2=0.9$ GeV$^2$ one gets a first glance at the 
expected falloff for the form factors mentioned previously.

The function $|G_T (q^2,W)|$, particularly how it evolves away from $q^2=0$,
has important consequences for the 
$N^\ast(1520) \to \gamma^\ast N$ transition.
In general, the $N^\ast \to \gamma^\ast N$  reactions 
are often studied under the assumption 
that the transition form factors in the timelike region
can be approximated 
by the experimental value of the form factors 
at the photon point ($q^2=0$), 
which implies that $W$ is fixed by 
the physical mass of the resonance
--- and therefore there is no $W$ dependence.
In the literature this approximation 
(no $q^2$ dependence of the electromagnetic coupling and $W=M_R$)
is known as the 
{\it QED approximation}, and it represents 
the form factor of a pointlike particle.
We also refer to this approximation 
as the constant form factor model. 
By construction, the constant form factor model
is not constrained by the form factor $G_C$,
because, at $q^2=0$, $G_C$ does not contribute to $|G_T|^2$,
according to Eq.~(\ref{eqGT}).
For finite $q^2$, however, $G_C$ contributes to $|G_T|^2$
with the term $\frac{q^2}{2W^2} |G_C|^2$.

%=============================================================
% corrected till here
%=============================================================

In Fig.~\ref{figGT} one can see that 
the value of $|G_T|$ at $q^2=0$ is close to 1,
consistently with the experimental value.
Therefore, in the constant form factor model,
$|G_T| \equiv 1.048$ underestimates 
the results from Fig.~\ref{figGT}.
Taking, for instance $q^2=0.9$ GeV, where $|G_T| \simeq 9$, 
one concludes that in the constant form factor model
$|G_T|$ is
about an order of magnitude too low.
Since the impact of $|G_T|$ 
in the decay widths is proportional to $|G_T|^2$,
in the range $q^2=0$--1 GeV$^2$
the constant form factor model may 
underestimate the decay widths
in 1 or 2 orders of magnitude.
For larger values of $q^2$, however, 
the discussion is different, 
since in our model, $|G_T|^2$ goes down with $q^2$ 
due to the multipole 
parametrizations of the meson cloud component.

%\newpage

The meson cloud dominance can be physically understood in terms
of the decay mechanisms of the resonances $N^\ast(1520)$ and $\Delta(1232)$.
The $\Delta(1232)$ decays almost exclusively into $\pi N$, but
the  $N^\ast(1520)$ can decay into $\pi N$ (40\%)
and into $\pi \pi N$ (60\%)~\cite{PDG}.
From the $\pi \pi N$ decays, one can expect then a stronger contribution
from the meson cloud effects for the transition form factors for the 
$N^\ast(1520)$ than for the $\Delta(1232)$.
The strength of these $\pi \pi N$ decays is encoded in the form factor data 
that we use in our fit to the physical spacelike form factor data.
In the  $\gamma^\ast N \to \Delta(1232)$ transition, 
the leading form factor is the magnetic form factor,
and in ours and other calculations, 
it is dominated by the valence quark contributions 
at low $q^2$~\cite{Timelike,Timelike2,NDelta,NDeltaD}.
There is, therefore, a smaller impact from the pion cloud.
By contrast, for the $\gamma^\ast N \to N^\ast(1520)$ transition, 
the electric and the magnetic form factors are both relevant at low $q^2$.
It is the different structure of form factors for the 
$\gamma^\ast N \to N^\ast(1520)$  and  
$\gamma^\ast N \to \Delta(1232)$ transition that implies 
a dominance of the  meson cloud effects, in the first case.

\subsection{Calculation of the decay width $\Gamma_{\gamma N}(W)$}

Using the formalism in Sec.~\ref{secDalitz},
we calculate the $N^\ast(1520) \to \gamma N$ decay width
as a function of $W$.
The $\gamma N$ decay width ($\Gamma_{\gamma N}$) is determined 
by the function $\Gamma_{\gamma^\ast N} (q,W)$, 
given by Eq.~(\ref{eqGamma1}) in the limit $q^2 =0$.
This width is then proportional 
to the function $|G_T (0,W)|^2$.

\begin{figure}[t]
\vspace{.3cm}
\centerline{
\mbox{
\includegraphics[width=2.8in]{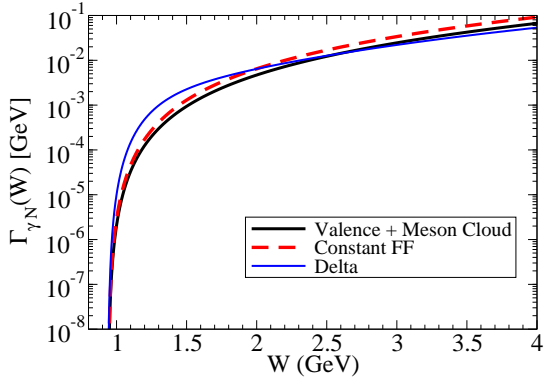}
}}
\caption{\footnotesize{
Decay width $\Gamma_{\gamma N}$ 
as function of $W$.
The current model (Valence + Meson Cloud) 
is compared with the constant form factor model 
and with the results obtained for 
the $\Delta(1232)$ (see Ref.~\cite{Timelike2}).
}}
\label{figGammaG}
\end{figure}

Our results for $\Gamma_{\gamma N}(W)$ 
are presented in Fig.~\ref{figGammaG} 
(thick solid line).
Since, as observed for the form factors, 
the dependence of  $|G_T |$ on $W$ is weak, 
the shape of $\Gamma_{\gamma N} (W)$ is determined 
mainly by the kinematic factor multiplying $|G_T |^2$ in  Eq.~(\ref{eqGamma1}).
The results are also compared with the
constant form factor model (dashed line). In the figure the results from 
the constant form factor for  $\Gamma_{\gamma N} (W)$
are close to the $q^2$ dependent results, but
we note that a logarithmic representation is used.
The deviation for large $W$ is about 30\%.
The similarity between the two results 
comes from the small dependence 
of our model on $W$ in the limit $q^2 \rightarrow 0$.
We expect that
for observables depending  on $q^2$
the results will differ much more. This is indeed the case, 
as confirmed in the next section.

To close this section, we compare  
 $\Gamma_{\gamma N} (W)$ with the 
$\Delta(1232) \to \gamma N$ decay width 
calculated in the Ref.~\cite{Timelike2}
also within the covariant spectator 
quark model framework (thin solid line). 
It is interesting that 
the $\Delta$ decay width is larger for small values of $W$, but the 
$N^\ast(1520)$ decay width is larger  when $W>2.5$ GeV.

\begin{figure}[t]
\vspace{.3cm}
\centerline{
\mbox{
\includegraphics[width=2.8in]{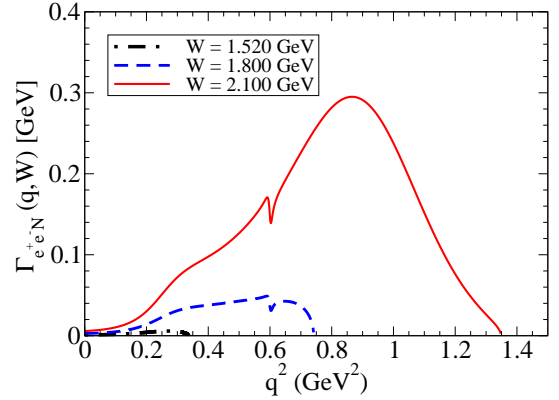}
}}
\caption{\footnotesize{
Dalitz decay width $\Gamma_{e^+ e^- N}(q^2,W)$  for 
$W=1.520, 1.800$ and 2.100 GeV.
}}
\label{figGammaD}
\end{figure}

\begin{figure}[t]
\vspace{.3cm}
\centerline{
\mbox{
\includegraphics[width=2.8in]{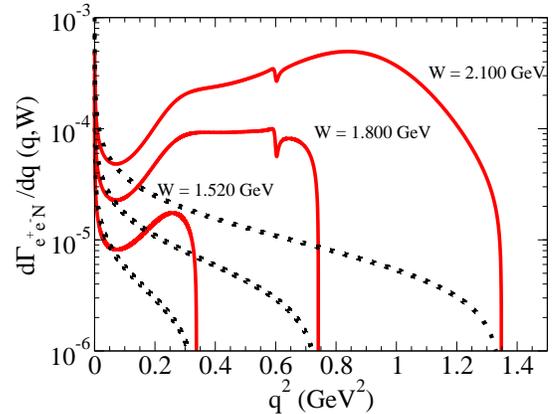}
}}
\caption{\footnotesize{ 
Results of  $\frac{d \Gamma}{d q} $ for  
$W=1.520, 1.800$ and 2.100 GeV (solid lines), compared 
to the estimate from 
the constant form factor model (dotted lines).}}
\label{figDGamma}
\end{figure}

\subsection{$N^\ast(1520)$ Dalitz decay}

We show now the results for the $N^\ast (1520)$ Dalitz decay,
$N^\ast (1520) \to e^+ e^- N$.
The Dalitz decay width
is determined by the function $\Gamma_{\gamma^\ast N} (q,W)$, 
given by Eq.~(\ref{eqGamma1}) 
for the case where $q^2$ is the 
photon momentum transfer  
for the dilepton decay $\gamma^\ast \to e^+ e^-$.

In Fig.~\ref{figGammaD},  we present 
the results of $\Gamma_{\gamma^\ast N} (q,W)$ 
for $W=1.52$, 1.8 and 2.1 GeV.
The dependence of $\Gamma_{\gamma^\ast N}$
on both variables, $q^2$ and $W$, is clear from the figure

Finally we show the dilepton decay rate $\frac{d \Gamma}{d q} (q,W)$,
where as before, we use the notation 
$\Gamma (q,W) \equiv \Gamma_{e^+ e^- N} (q,W)$.
The results for  $\frac{d \Gamma}{d q} $
for the three values of $W$ discussed previously 
are presented in Fig.~\ref{figDGamma}.
The results are compared with the constant form factor model.
The covariant spectator quark model 
differs significantly from the 
constant form factor model for $q^2 > 0.1$ GeV$^2$.
This effect is caused by the meson cloud contributions 
included in our model.

\begin{figure}[t]
\vspace{.3cm}
\centerline{
\mbox{
\includegraphics[width=2.8in]{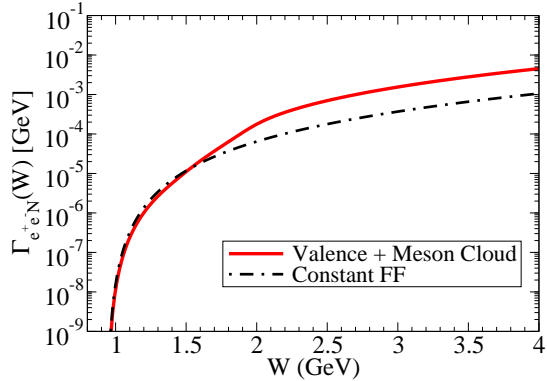}
}}
\caption{\footnotesize{
$N^\ast (1520)$ Dalitz decay width as a function of $W$.
The result of our model (solid line) is compared 
with the result of the constant form factor model
(dotted-dashed line).}}
\label{figDGammaDalitz0}
\end{figure}

\begin{figure}[t]
\vspace{.3cm}
\centerline{
\mbox{
\includegraphics[width=2.8in]{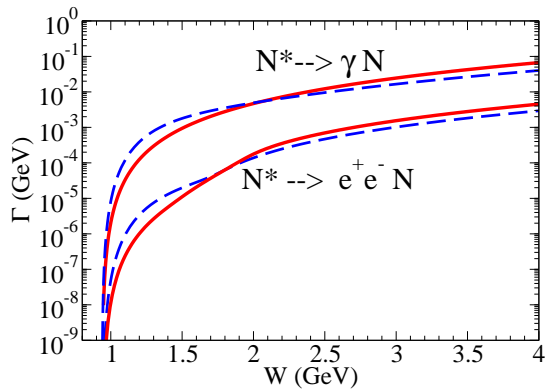}
}}
\caption{\footnotesize{
$N^\ast (1520)$ decay widths as function of $W$.
Photon and Dalitz decays (solid lines).
The results are also compared to 
the calculation for the $\Delta(1232)$ case 
(dashed lines).}}
\label{figDGammaDalitz}
\end{figure}

The function $\Gamma_{e^+ e^- N}(W)$ can now 
be evaluated integrating in $q$ 
according to Eq.~(\ref{eqGammaDal}).
The results are presented  in 
Fig.~\ref{figDGammaDalitz0}
that shows also the comparison with the results obtained
with the constant form factor model.
From the figure we can conclude 
that the effect of the $q^2$ dependence is diluted 
when we integrate in $q$ for $W < 1.6$ GeV.
One concludes then that  the $q^2$ dependence on 
$|G_T|^2$ is important 
when we look for the function $\Gamma_{e^+ e^- N}(q,W)$,
or for $\Gamma_{e^+ e^- N}(W)$ at a large $W$.

In  Fig.~\ref{figDGammaDalitz} 
we present the results of $\Gamma_{e^+ e^- N}(W)$ 
in comparison to the electromagnetic decay width 
$\Gamma_{\gamma N}(W)$.
In the same figure,
we compare also our results for 
$\Gamma_{e^+ e^- N}(W)$,  $\Gamma_{\gamma N}(W)$
from the $N^\ast(1520)$ resonance 
with the corresponding ones from 
the $\Delta(1232)$ decays~\cite{Timelike2}.
In both cases, one includes the combination 
of the valence quark and meson cloud contributions.

The results from  Fig.~\ref{figDGammaDalitz}  
imply that the two resonances are almost 
equally relevant for a large $W$,
suggesting that the $N^\ast(1520)$
may play an important role in dilepton 
decay reactions.

\section{Outlook and conclusions}
\label{secConclusions}

We apply to the $N^\ast(1520) \to \gamma N$ transition a model 
which adds a covariant valence quark core contribution 
with a meson cloud term.
The meson cloud term is related to
the pion electromagnetic form factor, 
which is well-established in the timelike region, and
the transition form factors are first  fixed in the spacelike region.
The form factor behavior in the timelike region
is then predicted, as well as the $N^\ast(1520) \to \gamma N$ 
decay width and the $N^\ast (1520)$ Dalitz decay,
$N^\ast (1520) \to e^+ e^- N$. 
The timelike $N^\ast(1520)$ transition form factors are dominated 
by the meson cloud contributions.

In the range $q^2=0$--1 GeV$^2$ 
the constant form factor model or {\it QED approximation}
that is usually taken in the literature
underestimates the electromagnetic coupling of the $N^\ast(1520)$
up to 2 orders of magnitude. 
This has a large effect on $q^2$ dependent observables 
as the $N^\ast (1520)$ Dalitz decay.
The $q^2$ dependence effect   
may be diluted in $\Gamma_{e^+ e^- N}(W)$ which is
obtained by integrating over $q^2$, but 
it can be clearly observed if we 
look at the differential Dalitz decay width $\frac{d \Gamma}{dq} (q,W)$. 

In line with the HADES results~\cite{HADES14,Ramstein16a,Przygoda16a}, 
the $N^\ast(1520)$ and $\Delta(1232)$  
decays compete, 
and at large energies the first is certainly important.

\begin{acknowledgments}
G.R.~was supported by the Brazilian Ministry of Science,
Technology and Innovation (MCTI-Brazil). 
M.T.P.~received financial support from Funda\c{c}\~ao 
para a Ci\^encia e a Tecnologia (FCT) under Grants 
Nos.~PTDC/FIS/113940/2009, CFTP-FCT 
(PEst-OE/FIS/U/0777/2013) and POCTI/ISFL/2/275. 
This work was also partially supported by the European Union 
under the HadronPhysics3 Grant No.~283286. 
\end{acknowledgments}

\appendix

\section{Regularization of high momentum poles}

As discussed in the main text, for a given $W$
the squared momentum $q^2$ is limited by the condition 
$q^2 \le (W-M)^2$.
Then, if one has a singularity for $q^2=\Lambda^2$,
that singularity will appear for values of $W$ such 
that $\Lambda^2 \le (W-M)^2$, or $W \ge M + \Lambda$.

To avoid a singularity at $q^2=\Lambda^2$, where $\Lambda^2$
is one of the cutoffs introduced in our meson cloud 
parametrizations and in the quark current (pole at $q^2=M_h$)
we will use a simple procedure.
We start with 
\ba
\frac{\Lambda^2}{\Lambda^2 - q^2} 
\to \frac{\Lambda^2}{\Lambda^2 - q^2 - i \Lambda \Gamma_X(q^2)},
\label{eqR1}
\ea
where 
\ba
\Gamma_X(q^2)= 4 \Gamma_X^0
\left( \frac{q^2}{q^2+ \Lambda^2}
\right)^2
\theta(q^2), 
\label{eqGammaX}
\ea
In the last equation $\Gamma_X^0$ is a constant 
given by  $\Gamma_X^0 = 4 \Gamma_\rho^0\simeq 0.6$ GeV.

The function $\Gamma_X(q^2)$ defined by 
Eq.~(\ref{eqGammaX}) is  $\Gamma_X(q^2)=0$ 
when $q^2 < 0$, and continuously extended for $q^2 > 0$.
Therefore the results in spacelike (where there are no singularities) 
are kept unchanged. 
The factor $4 \Gamma_X^0$
was chose in order to obtain $\Gamma_X = \Gamma_X^0$, 
for $q^2= \Lambda^2$, and  
$\Gamma_X \simeq 4  \Gamma_X^0$ 
for very large $q^2$.
Finally the value of   $\Gamma_X^0$ was chosen 
to avoid very narrow peaks around $\Lambda^2$.
 
Contrarily to the width $\Gamma_\rho(q^2)$ associated 
with the $\rho$-meson pole in the quark current
which has nonzero values only when $q^2 > 4 m_\pi^2$,
one has for $\Gamma_X(q^2)$ nonzero values 
also in the interval $4 m_\pi^2 > q^2 >0$.
However, the function $\Gamma_X(q^2)$ varies smoothly 
in that interval and its values are very small
when compared to $\Gamma_X^0$.

The procedure given by Eqs.~(\ref{eqR1})-(\ref{eqGammaX})
was used already in applications 
of the model from Ref.~\cite{Timelike} 
in the calculation of the $\gamma^\ast N \to \Delta$ 
form factors in the timelike regime~\cite{Timelike2,Weil12}.
With this procedure the emerging singularities  
for $W > 2.17$ GeV are avoided and the results 
are almost identical to the 
results for $W < 2.17$ GeV, without regularization.
The singularity that appears 
for $W \simeq 2.17$ GeV is due to the pion cloud parametrization
 of $G_M$~\cite{Timelike}.

As in most cases,the high momentum effects and the high $q^2$ 
contributions are suppressed, and the details of the
regularization procedure are not important, 
as far as removing the spurious singularities is concerned.

For the tripole factors associated with the functions
(\ref{eqG4pi})-(\ref{eqGCpi}) we use
\ba
\left( \frac{\Lambda^2}{\Lambda^2 -q^2}
\right)^3 \to
\left( 
\frac{\Lambda^4}{(\Lambda^2 -q^2)^2 + \Lambda^2 [\Gamma_X(q^2)]^2}
\right)^{3/2},
\ea
where $\Gamma_X(q^2)$ is obtained by Eq.~(\ref{eqGammaX}).

%\clearpage

%\input{biblo}

\end{document}